\documentclass[%
 reprint,
superscriptaddress,
 amsmath,amssymb,
 aps,
pra,
]{revtex4-2}
\usepackage[utf8]{inputenc}
\usepackage{graphicx}
\usepackage{dcolumn}
\usepackage{bm}
\usepackage{hyperref}
\usepackage[normalem]{ulem}

\usepackage{upgreek}
\usepackage{xcolor}
\usepackage[T1]{fontenc}

\newcommand{\ket}[1]{{\left | #1 \right >}}

\begin{document}


\title{Intercombination line photoassociation spectroscopy of ${}^{87}$Rb${}^{170}$Yb}

\author{Tobias Franzen}\email{tobias.franzen@durham.ac.uk} \altaffiliation[Present address: ]{Joint Quantum Centre (JQC) Durham-Newcastle, Department of Physics, Durham University, South Road, Durham, DH1 3LE, United Kingdom.}
\affiliation{Institut für Experimentalphysik, Heinrich-Heine Universität Düsseldorf, 40225 Düsseldorf, Germany}

\author{Bastian Pollklesener}
\affiliation{Institut für Experimentalphysik, Heinrich-Heine Universität Düsseldorf, 40225 Düsseldorf, Germany}
 \author{Christian Sillus}
\affiliation{Institut für Experimentalphysik, Heinrich-Heine Universität Düsseldorf, 40225 Düsseldorf, Germany}
\author{Axel Görlitz}\email{axel.goerlitz@hhu.de}
\affiliation{Institut für Experimentalphysik, Heinrich-Heine Universität Düsseldorf, 40225 Düsseldorf, Germany}

\date{\today}

\begin{abstract}
We report on the first observation of photoassociation (PA) near the $\text{[Rb]}\, 5\text{s}\, {}^2 S _\frac 1 2 + \text{[Yb]}\, 6\text{s}\,6\text{p}\, {}^3P_1$-asymptote in a mixture of ${}^{87}$Rb and ${}^{170}$Yb. In a search spanning binding energies between 0.1\,GHz and 11\,GHz, a single pair of interspecies PA resonances is detected around 3.1\,GHz. These resonances are characterized by extracting PA rates, binding energies and Zeeman shift coefficients.
Using one of these resonances, 2-photon-photoassociaton is performed, improving on previous measurements of the binding energies of the two least bound states in the electronic ground state and demonstrating intercombination line photoassociation as a powerful spectroscopic tool. We discuss implications for pathways towards RbYb molecules in the absolute ground state.
\end{abstract}

\maketitle


\section{Introduction}
The production of ultracold and ultimately quantum degenerate samples of polar molecules has recently attracted much attention, with applications in quantum simulation \cite{micheli2006toolbox,georgescu14quantum,Gadway_2016,daley2022practical}, quantum chemistry  \cite{krems08cold,balakrishnan2016perspective,liu2022bimolecular} and quantum computing \cite{gregory2021robust,sawant20qudits}.
Molecules with a ${}^2\Sigma$ ground state, such as dimers composed of an alkali and a closed-shell atom promise to further extend the rich internal structure of these systems and provide an additional degree of freedom due to their magnetic dipole moment \cite{micheli2006toolbox,Quemener2016,Abrahamsson2007,Perez-Rios2010,Herrera2014,Karra2016,Asnaashari_2022}.

Despite considerable advances \cite{fitch2021laser,mccarron2018magnetic,cheuk2018lambda,anderegg2019optical,truppe2017molecules,williams2018magnetic}, direct laser cooling of molecules remains extremely challenging.
An alternative route that has been applied to bi-alkalis \cite{Ni2008,Lang2008,Danzl2008,Molony2014,Park2015,Seeselberg2018a} and the homonuclear closed-shell molecule Sr\textsubscript{2} \cite{leung2021ultracold} is the creation of ultracold molecules from ultracold atoms.
In particular in bialkalis great progress has been made recently, with several groups producing degenerate samples \cite{cao2022preparation,de2019degenerate,schindewolf2022evaporation}.

Magnetic Feshbach resonances \cite{Chin2010} are the association mechanism of choice in bialkalis, where the interaction between electron spins provides strong coupling between channels. However, the picture is less clear for a combination of an alkali and a closed-shell atom, where the \textsuperscript{1}S\textsubscript{0} ground state means that these couplings are absent. 
While Feshbach resonances due to distance-dependent hyperfine couplings have been predicted \cite{Zuchowski2010,brue13prospects,Brue2012} and observed in Rb + Sr \cite{Barbe2018}, Li + Yb \cite{Green2020} and Cs + Yb \cite{franzen2022observation}, these resonances are extremely narrow and their use for magnetoassociation remains an outstanding challenge. Magnetic Feshbach resonances in Rb + Yb have been predicted \cite{brue13prospects} but not observed to date.

On the other hand, efficient all-optical production of Sr\textsubscript{2} molecules in the absolute ground state has recently been reported, demonstrating the feasibility of this approach. In that work narrow line photoassociation was followed by spontaneous decay to a weakly bound level in the electronic ground state, from where the absolute ground state was reached by \textit{stimulated Raman adiabatic passage} (STIRAP) \cite{leung2021ultracold}.
Previous studies, also in Sr\textsubscript{2}, have also demonstrated efficient free-bound STIRAP \cite{ciamei2017efficient}, suggesting the possibility of a fully coherent transfer.  
Photoassociation thus remains a promising alternative technique for the creation of alkali--closed-shell molecules.

The combination Rb + Yb is an attractive candidate as cooling of both species to degeneracy is well established and the large number of stable isotopes in Yb allows the creation of both fermonic and bosonic molecules with a range of reduced masses.
Previous photoassociation experiments in RbYb \cite{nemitz2009production,Bruni2016hyperfine,munchow2011two,Borkowski2013rbyb} have used the ${}^2\Pi_\frac 1 2$ potential corresponding to the atomic threshold $\text{[Rb]}\,5\text{p}\,{}^2 P _\frac 1 2 + \text{[Yb]} \, 6\text{s}^{2} \, {}^1S_0$, where the strong D1 line provides large photoassociation rates.
These investigations included the measurement of the differential hyperfine shift in the ${}^2\Pi_\frac 1 2$ potential \cite{Bruni2016hyperfine}, analogous to the shift responsible for some of the Feshbach resonances in the ground state. In addition the ground state potential was characterized by 2-photon-photoassociation \cite{munchow2011two}, leading to the accurate determination of the scattering lengths for all isotope combinations \cite{Borkowski2013rbyb}.

\begin{figure}[tb]
\includegraphics[width=0.9\linewidth]{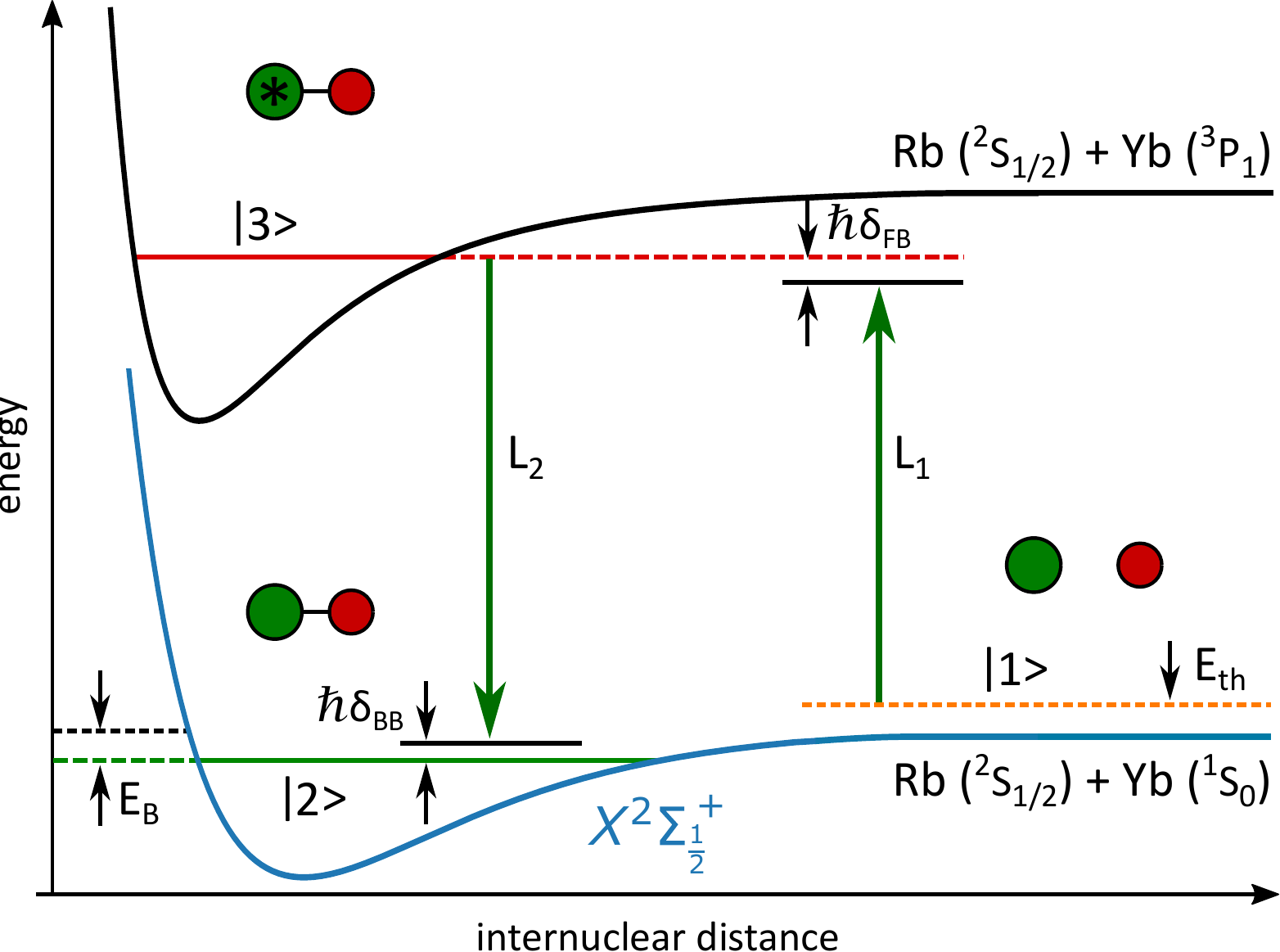}
\caption{\label{fig:paintro} Schematic diagram of the relevant ground and excited state potentials of RbYb and the free-bound (L$_1$) and bound-bound (L$_2$) transitions used in this work. } 
\end{figure}

A unique feature of closed-shell atoms is the presence of narrow intercombination line transitions. In Yb the transitions to the ${}^3P_j$ manifold exhibit natural linewidths ranging from the mHz range for the ${}^1S_0 \rightarrow {}^3P_0$ \textit{clock transition} \cite{porsev04clock} to $2\pi\cdot181\,\mathrm{kHz}$ for the ${}^1S_0 \rightarrow {}^3P_1$ transition, which is commonly used for narrow-line laser cooling \cite{kuwamoto96magnetooptical}. While the former provides an extremely sensitive spectroscopic probe allowing for example the measurement of on-site interactions in  optical lattices \cite{scazza2014observation,Franchi_2017}, photoassociation would be extremely challenging due to the small transition dipole moments. Instead we chose to work near the ${}^1S_0 \rightarrow {}^3P_1$ transition, as illustrated in Fig.~\ref{fig:paintro}. With a  transition linewidth about thirty times smaller than that of the D1 line of Rb, it allows for spectroscopy of more weakly bound states with significantly higher resolution, while still providing sufficient transition strength to locate resonances without prior knowledge of binding energies.
The $\text{[Rb]}\, 5\text{s}\, {}^2 S _\frac 1 2 + \text{[Yb]}\, 6\text{s}\,6\text{p}\, {}^3P_1$-asymptote connects to three potentials labeled by ,$^{(2S+1)}\Lambda_{\Omega} =$ $1^{4}\Pi_{\frac{1}{2}}$, $1^{4}\Pi_{-\frac{1}{2}}$, $2^{2}\Pi_{\frac{3}{2}}$ in the short range with Hund's case (a) and $(n)\Omega=$ $(5)\frac{1}{2}$, $(1) -\frac{1}{2}$, $(2) \frac{3}{2}$ in the long range with Hund's case (c) \cite{shundalau2017abinitio}. Prior to this work no experimental data on these potentials was available.

Further motivation to explore bound states of the ${}^3P_j$ manifold is given by the recent prediction of Feshbach resonances between Rb and metastable Yb in the $j=0,2$ states \cite{mukherjee2021feshbach,mukherjee2022magnetic}.

This paper is organized as follows: In Sec.~\ref{sec:expsetup} we briefly describe our apparatus and the preparation sequence used for the measurements reported here. Our observations of intercombination line photoassociation resonances and their characterization are presented in Sec.~\ref{sec:1photpa}. Finally, in Sec.~\ref{sec:2photpa}, we describe 2-photon photoassociation using this transition and precisely determine the binding energy of the two least bound states of the electronic ground state.

\section{Experimental setup}\label{sec:expsetup}

In our new apparatus (described in detail elsewhere \cite{apparatus}), Rb and Yb are prepared in separate vacuum chambers by magneto-optical trapping and microwave driven evaporation in a quadrupole trap for Rb and an intercombination line magneto-optical trap (MOT) for Yb. Both species are then loaded into single beam optical dipole traps and subsequently transported to a common science chamber by axial translation of these traps. Here further evaporative cooling and merging of the two samples takes place in crossed optical dipole traps. All dipole traps operate close to 1064\,nm, leading to trap depths that are significantly higher for Rb than for Yb. This in turn leads to a similar difference in temperature.

We chose the combination of ${}^{87}$Rb + ${}^{170}$Yb due to its favorable intra- and interspecies scattering lengths of $a_\mathrm{Rb-Rb} = 100\,a_0$ \cite{Mertes07Nonequilibrium}, $a_\mathrm{Yb-Yb} = 64\,a_0$ \cite{Kitagawa2008twocolor} and $a_\mathrm{Rb-Yb} = -11\,a_0$ \cite{Borkowski2013rbyb}. These values ensure efficient preparation of each species by evaporative cooling in optical dipole traps as well as miscibility. The previous observation of phase separation well before the onset of degeneracy in mixtures of ${}^{87}$Rb + ${}^{174}$Yb \cite{Baumer2011spatial} demonstrates this can not be taken for granted even in thermal samples. 

 In a typical experiment we prepare $3\cdot10^5$ ${}^{170}$Yb atoms at a temperature of $1.5\,\mathrm{\upmu K}$ and $2\cdot10^5$ Rb atoms at a temperature of $10\,\mathrm{\upmu K}$ in a crossed optical dipole trap.
The calculated trap frequencies along the principal directions of the trap are $\omega_\mathrm{Yb} = 2\pi \cdot ( 80\,\mathrm{Hz}, 250\,\mathrm{Hz}, 260\,\mathrm{Hz})$ and $\omega_\mathrm{Rb} = 2\pi \cdot ( 0.35\,\mathrm{kHz}, 1.1\mathrm{kHz}, 1.1\,\mathrm{kHz})$, resulting in estimated peak densities of $n_\mathrm{Yb} \approx 1\cdot10^{13}\,\mathrm{cm}^{-3}$ and  $n_\mathrm{Rb} \approx 2\cdot10^{13}\,\mathrm{cm}^{-3}$, respectively, with phase space densities of $\sim10^{-2}$ for both species. Due to the small interspecies scattering cross section no appreciable interspecies thermalization, which with our current trapping potentials would lead to the loss of Yb, is observed. Rb is prepared almost exclusively in the $(f=1, m_f=-1)$ hyperfine state. 

Photoassociation light is applied for typically $1\,\mathrm{s}$ after which the remaing atom number is detected. Despite similar atom numbers before photoasociation, rapid loss of Yb from off-resonant excitation near the atomic line limits the contrast observed in the Rb atom number and the resonances are more easily detected in the Yb atom number. Light to drive the free-bound and bound-bound transitions is generated using a frequency doubled fiber laser system, frequency stabilized at a tunable offset from the MOT laser system (see Appendix \ref{sec:laserstab}). Light to address the bound-bound transition is derived from the same light source (see Appendix \ref{sec:boundbound}). 

\section{1-photon photoassociation}
\label{sec:1photpa}

The narrow linewidth of the atomic transition and the lack of prior information make the search for intercombination line PA resonances challenging.
Ab-initio calculations predict the $C_6$ coefficient \cite{porsev2014relativistic}, but not the fractional part of the dissociation quantum number $\nu_D$ and can thus be used to estimate the spacing of bound levels, but not their actual positions in a given isotopologue. In a Hund's case (c) picture the relevant potential energy curves have either $\left | \Omega \right | = \frac 1 2$ or $\left | \Omega \right | = \frac 3 2$

\begin{figure}[tb]
\includegraphics[width=0.8\linewidth]{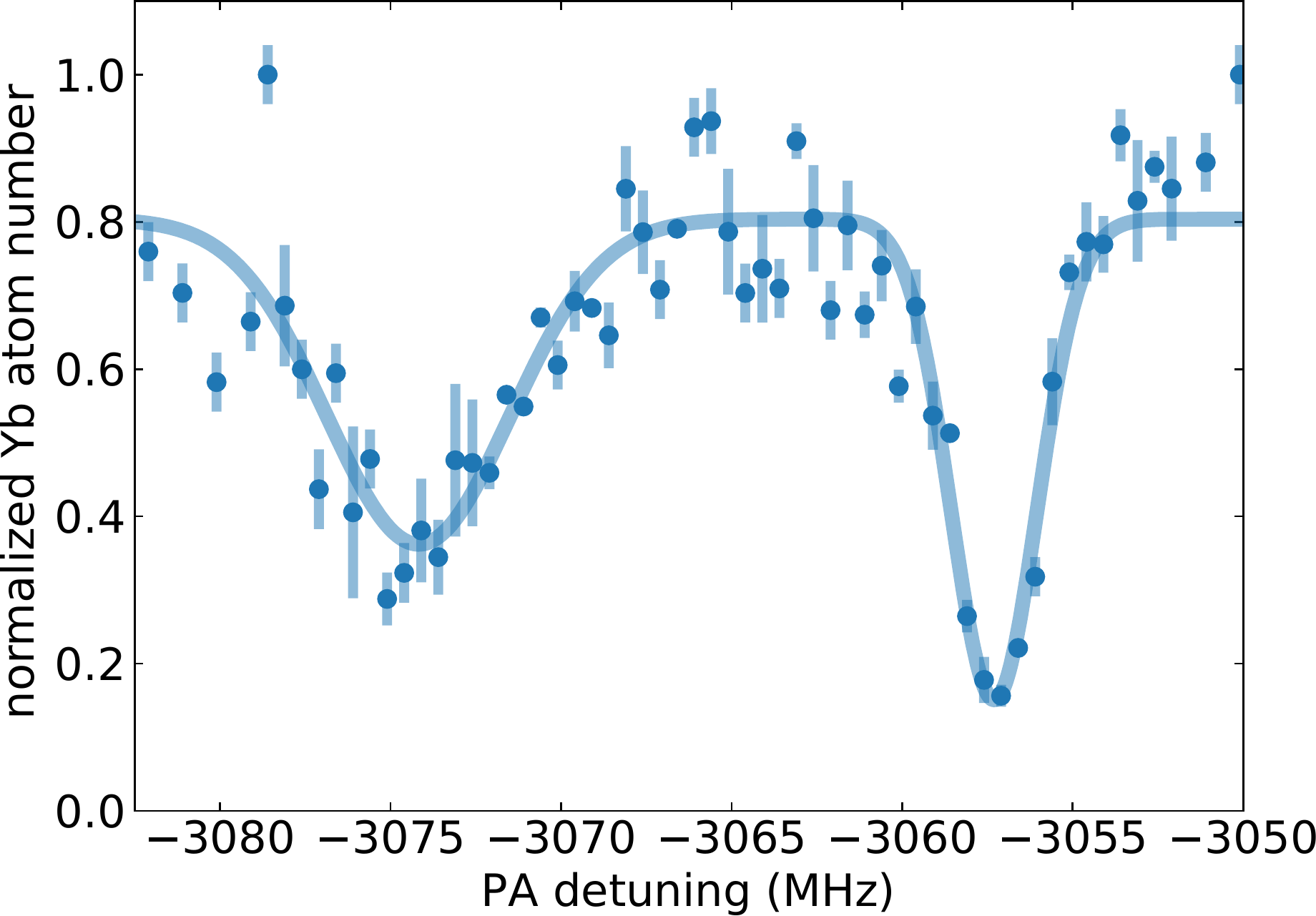}
\caption{\label{fig:paonoff} Free-bound photoassociation lines observed in the Yb atomnumber. The solid line is a Gaussian fit as guide to the eye. }
\end{figure}

We perform photoassociation spectroscopy at detunings between $0.1\,\mathrm{GHz}$ and $11\,\mathrm{GHz}$, locating a single pair of resonances around $3.1\,\mathrm{GHz}$, shown in Fig. \ref{fig:paonoff}. In measurements in the absence of Rb these features disappear, ruling out the (remote) possibility of an intra-species photoassociation resonance.

\begin{figure}[b]
\includegraphics[width=\linewidth]{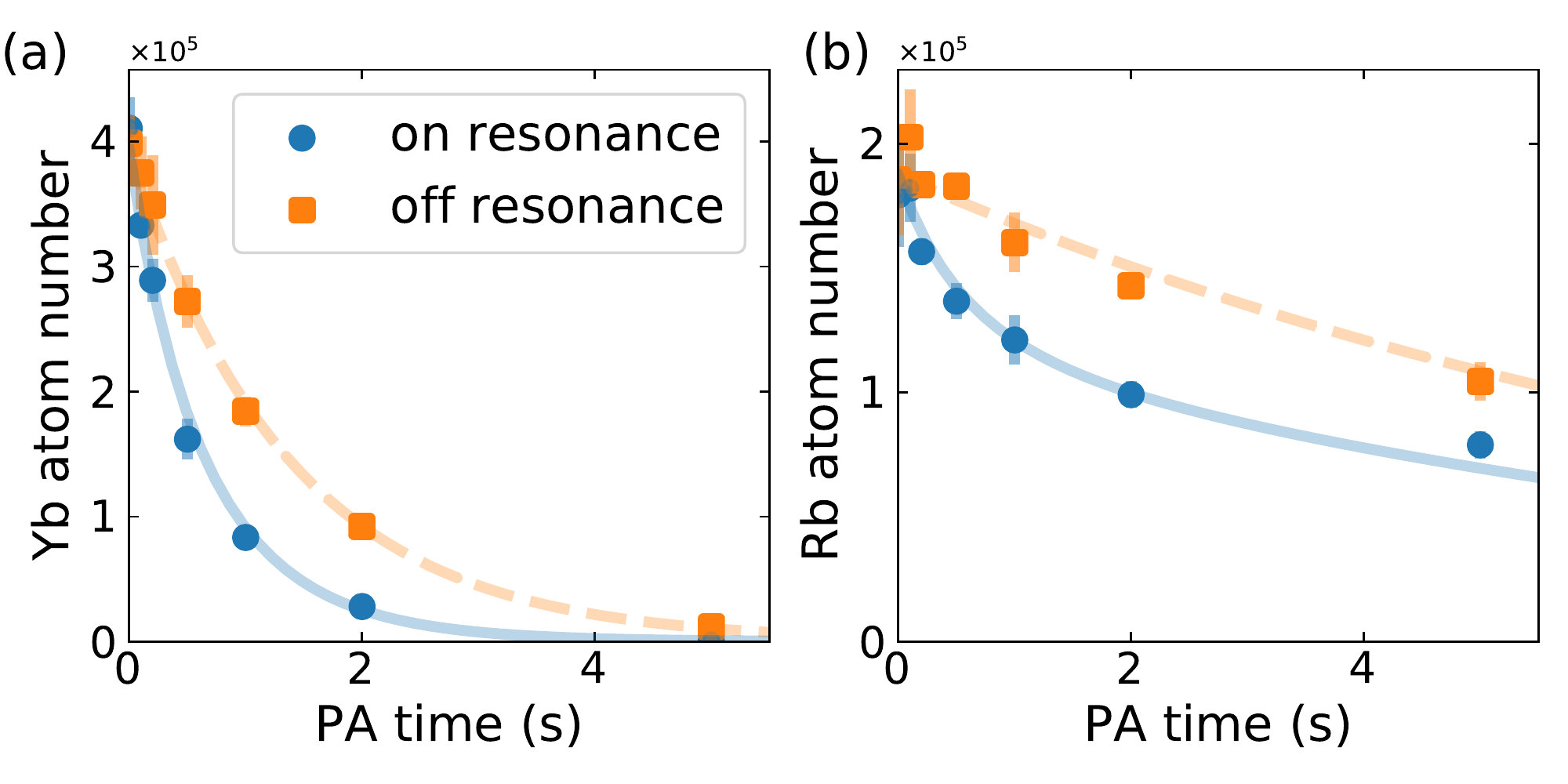}
\caption{\label{fig:decay} Decay curves for (a) Yb and (b) Rb with the PA light on and off resonance for the stronger transition located at $-3057\,\text{MHz}$}.
\end{figure}

To quantify the strength of the PA line we measure the time dependence of the atom number for each species with the PA laser tuned on and off resonance as shown in Fig.~\ref{fig:decay}. We fit the data with solutions of a simple rate equation model for the atomic densities $n_\mathrm{Rb}$ and $n_\mathrm{Yb}$ given by
\begin{align}\label{eq:parateode}
    \dot n_\mathrm{Rb} & = - n_\mathrm{Rb}\cdot K_\mathrm{Rb}  -  n_\mathrm{Rb}\cdot n_\mathrm{Yb}\cdot K_\mathrm{PA} \\
    \dot n_\mathrm{Yb} & = - n_\mathrm{Yb}\cdot K_\mathrm{Yb}  -  n_\mathrm{Rb}\cdot n_\mathrm{Yb}\cdot K_\mathrm{PA} 
\end{align}
with the photoassociation rate constant $K_\mathrm{PA}$ and the single species loss constants $K_\mathrm{Rb},K_\mathrm{Yb}$ \footnote{Inclusion of three-body losses does not improve the fit due to the relatively low Rb density at this stage of the experiment.}.
From this fit we extract a photoassociation rate constant of $K_\mathrm{PA} \approx 5\cdot10^{-14}\,\mathrm{cm}^3/\mathrm{s}$ for the slightly stronger resonance at $-3057\,\mathrm{MHz}$ for a PA laser intensity of $\sim 40\,\text{W/cm$^2$}$. 
Even considering the difference in atomic linewidths, this is significantly larger than predictions made for RbSr in \cite{devolder2018proposal}, but nonetheless orders of magnitude smaller than typical PA rate constants obtained near the alkali D-line asymptotes. 
The single species loss constant of $0.7\,\mathrm{s}^{-1}$ for Yb is an order of magnitude larger than for Rb \footnote{We attribute the relatively large loss rate for Rb to heating from the multi mode trapping laser.} which is explained by off-resonant scattering. While the PA rate is not saturated and could be increased with higher laser intensity, this would also lead to a corresponding increase in the Yb single species loss.

Despite extensive efforts we could not locate any other resonances.  For most of the search range, transitions with a PA rate constant $K_\mathrm{PA} \gtrsim 2\cdot10^{-14}\,\mathrm{cm}^3/\mathrm{s}$ would have been detected with a high probability. 
Sensitivity is however strongly degraded for binding energies less than $\sim h\cdot2\,\mathrm{GHz}$ due to increased off-resonant excitation of the atomic transition and in the immediate vicinity of Yb\textsubscript{2} resonances, both  requiring a reduction in photoassociation pulse area by several orders of magnitude. We note that with the observed resonance position and the $C_6$-coefficients predicted in Ref.~\onlinecite{porsev2014relativistic}, the two neighbouring PA resonances are indeed expected to be close to Yb\textsubscript{2} resonances.
Sensitivity would also be degraded by a factor of up to four \footnote{Assuming their linewidth is no smaller than the atomic linewidth of $2\,\pi\cdot 182\,\mathrm{kHz}$} for transitions with an effective linewidth of less than the typical step size of $2\,\pi\cdot 1\,\mathrm{MHz}$.

\begin{figure}[b]
\includegraphics[width=\linewidth]{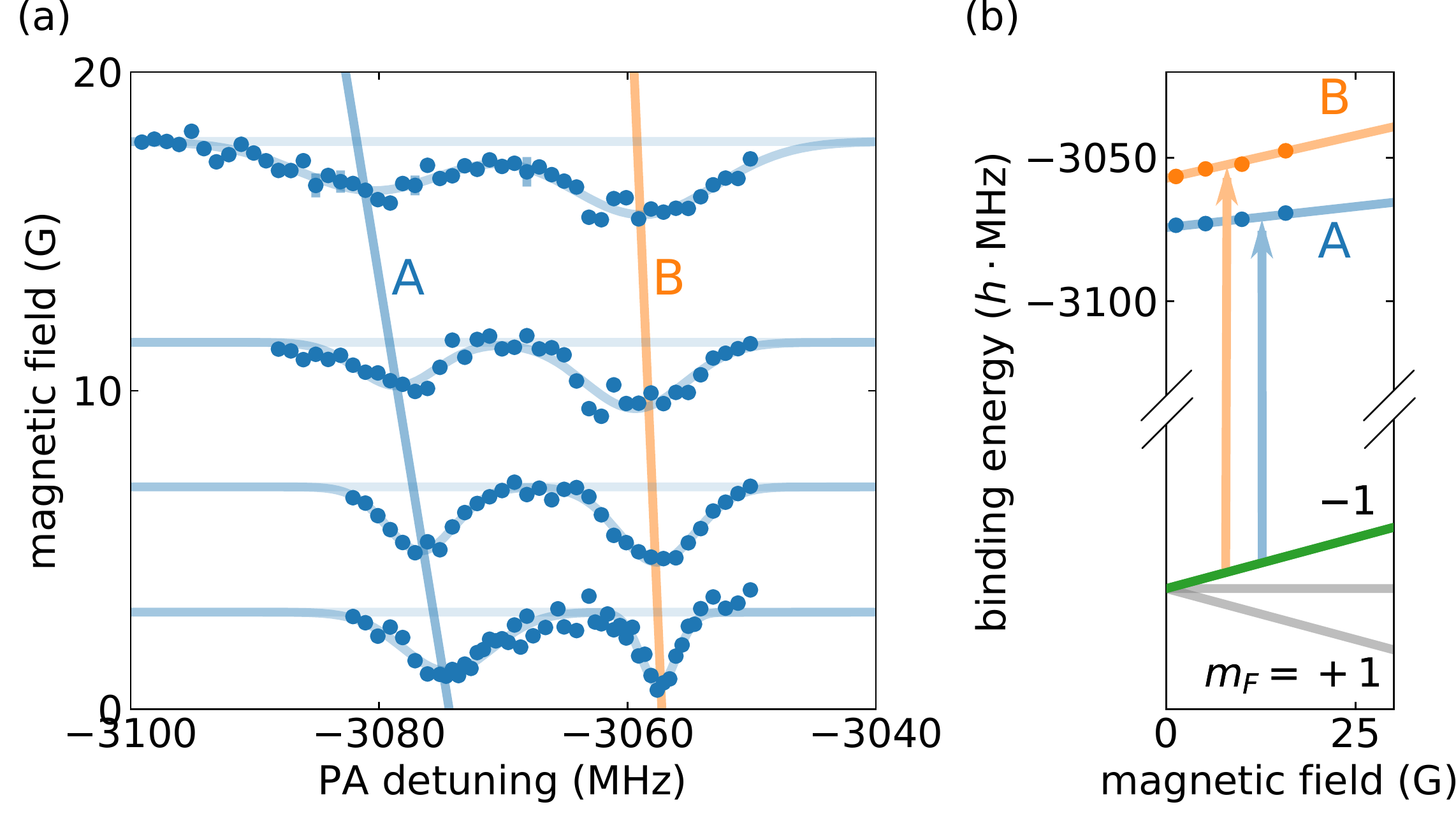}
\caption{\label{fig:zeeman} Investigation of the Zeeman shift of the observed transitions. \textbf{(a)} Spectra at various magnetic fields, offset by the magnetic field strength. \textbf{(b)} reconstructed shift of A and B from Gaussian fits shown in (a)}
\end{figure}

To further characterize the observed resonances, we measure their Zeeman shift as shown in Fig.~\ref{fig:zeeman}.
The magnetic field values were calibrated using microwave spectroscopy in Rb. The polarization was undefined for this measurement, so potentially both $\pi$ and $\sigma_{\pm}$ transitions contribute to the spectrum. Additionally, residual background field contributions may cause a shift in magnetic field orientation at low fields.

Both loss features are fit by Gaussians and their center positions, taking into account the Zeeman shift of the atomic threshold, are shown in Fig.~\ref{fig:zeeman}b.
The fits shown yield shift coefficients of $0.6(1)\,\mathrm{MHz/G}$ and $0.3(1)\,\mathrm{MHz/G}$ respectively, with zero-field transition frequencies of $-3057.2(3)\,\mathrm{MHz}$ and $-3074.3(3)\,\mathrm{MHz}$.  Thus the two peaks obviously remain split even at zero magnetic field. Yet their close proximity and similar strength suggest that they are in fact closely related. 

Given the small binding energies, it appears reasonable to treat the molecular states as combinations of the atomic states.
The shift coefficients for the Rb states in $f=1$ are $-m_f\cdot 0.7\,\mathrm{MHz/G}$ , while the excited \textsuperscript{3}P\textsubscript{1} state of Yb  shifts with $m_j\cdot 2.1\,\mathrm{MHz/G}$. All combinations of these states will thus shift with multiples of $0.7\,\mathrm{MHz/G}$ and cannot satisfactorily explain the observed shifts. Our efforts to explain the Zeeman shift in the frame work of \textit{Hund's case (c)} or \textit{(e)} have also failed to deliver a consistent assignment.  

In addition to the shifts, Fig.~\ref{fig:zeeman} shows a significant broadening of the peaks with increasing magnetic field. 
For the zero-field full-width half-maximum (FWHM) of the peak at $-3057.2(3)\,\mathrm{MHz}$ a value of $3\,\mathrm{MHz}$ is extrapolated. Even this linewidth is still an order of magnitude larger than the atomic linewidth. No further reduction of the linewidth was observed when reducing the PA intensity, ruling out saturation broadening. We rule out an increase in magnetic field noise for higher fields as the cause for the broadening by microwave spectroscopy on Rb, which is also used for the calibration of the field.

Significant broadening of the intercombination line photoassociation resonances has previously been observed in RbSr \cite{ciamei18taming}. In that work it was argued that this could be caused by strong radiative decay to bound states in the electronic ground state. 
In contrast to the alkali D-lines, the ${}^1S_0 \rightarrow {}^3P_1$ is only weakly allowed by mixing of the ${}^3P_1$ and ${}^1P_1$ states. At shorter internuclear distances, the presence of the Rb atom leads to further mixing, increasing the  transition dipole moment \cite{shundalau2017abinitio}.
However, from simple estimates based on model potentials and the transition dipole moment curves calculated by Shundalau and Minko \cite{shundalau2017abinitio} we expect the increase in transition dipole moment for this state to be less than $10\,\%$ of the atomic dipole moment. This mechanism thus seems unlikely to significantly affect the state lifetime or transition rates in the case of RbYb.

Another possible explanation for the observed linewidths is non-radiative decay within the \textsuperscript{3}P manifold, for example through predissociation to the \textsuperscript{3}P\textsubscript{0} state. Such decay has for example been encountered in the analysis of Feshbach resonances associated with the \textsuperscript{3}P\textsubscript{2} state \cite{mukherjee2021feshbach}, where it leads to strongly decayed resonances, in which the closed channel lifetime is limited to the order of microseconds by predissociation to the \textsuperscript{3}P\textsubscript{0} continuum.

Interestingly, the calculations performed for Feshbach resonances in the \textsuperscript{3}P\textsubscript{2} state also show a decay width that is increasing with magnetic field in the same range that is covered by our measurements \cite{mukherjee2021feshbach}. If a similar effect occurs for the \textsuperscript{3}P\textsubscript{1} state, this could explain the observed behaviour of the linewidths.

 A somewhat similar situation has also been described for the ${}^3\mathrm{P}_j$ manifold of the molecular ion HeN\textsuperscript{+}  \cite{soldan2002near}. Notably this leads to many transitions being broadened -- in particular in the presence of a magnetic field -- to an extent that makes them unobservable and might provide an explanation for the fact that we have only observed a single pair of seemingly related transitions. 
A definitive assignment of the observed  lines and thus the observed Zeeman shifts and linewidths will require further experimental and theoretical work beyond the scope of this paper.

Taking into account the effects of Zeeman and thermal shift, the binding energies of the two states -- assuming they can be assigned to the $f=1$ threshold of Rb --  are measured to be
$3057.0(5)\,\mathrm{MHz}$ and $3074.1(5)\,\mathrm{MHz}$ and are thus split by $17.1(7)\,\mathrm{MHz}$.
Taking the $C_6$-values calculated by Porsev et al \cite{porsev2014relativistic}, these binding energies correspond to a classical outer turning point of $ 45\,a_0$ to $ 46\,a_0$ for $\left |\Omega \right | = \frac 1 2$ and $\left |\Omega \right | = \frac 3 2$, respectively.
This is at significantly longer range than the most weakly bound level that was observed for photoassociation near the D1 line of Rb with an outer turning point around $35\,a_0$ \cite{nemitz2009production,Bruni2016hyperfine}. This is due to the difference in $C_6$-coefficients and the lower linewidth allowing for spectroscopy closer to the asymptote.
In combination with the predicted $C_6$-coefficients, the neighbouring levels may be predicted to have binding energies around $1.2\,\mathrm{GHz}$ and $6\,\mathrm{GHz}$, respectively, though with large uncertainties. Both of these are however close to Yb\textsubscript{2} resonances which could obscure the RbYb resonances.

The low photoassociation rate and the significant off-resonant light scattering on the atomic transition will make the use of the observed transitions for the efficient production of RbYb molecules challenging. However, intercombination line photoassociation presents a powerful tool for the characterization of weakly bound states in both the electronically excited and the electronic ground state.

\section{2-photon photoassociation}\label{sec:2photpa}

\begin{figure}[b]
(a)\hspace{.9\linewidth}~\\
\includegraphics[width=\linewidth]{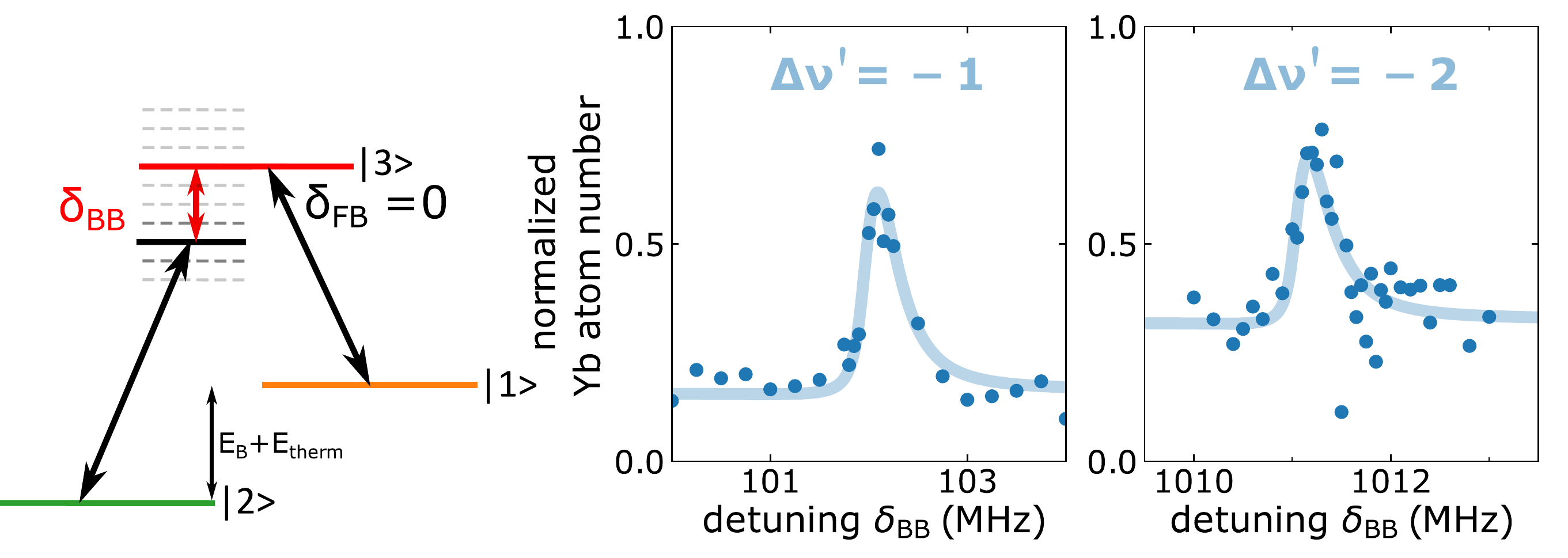}
(b)\hspace{.9\linewidth}~\\
\includegraphics[width=\linewidth]{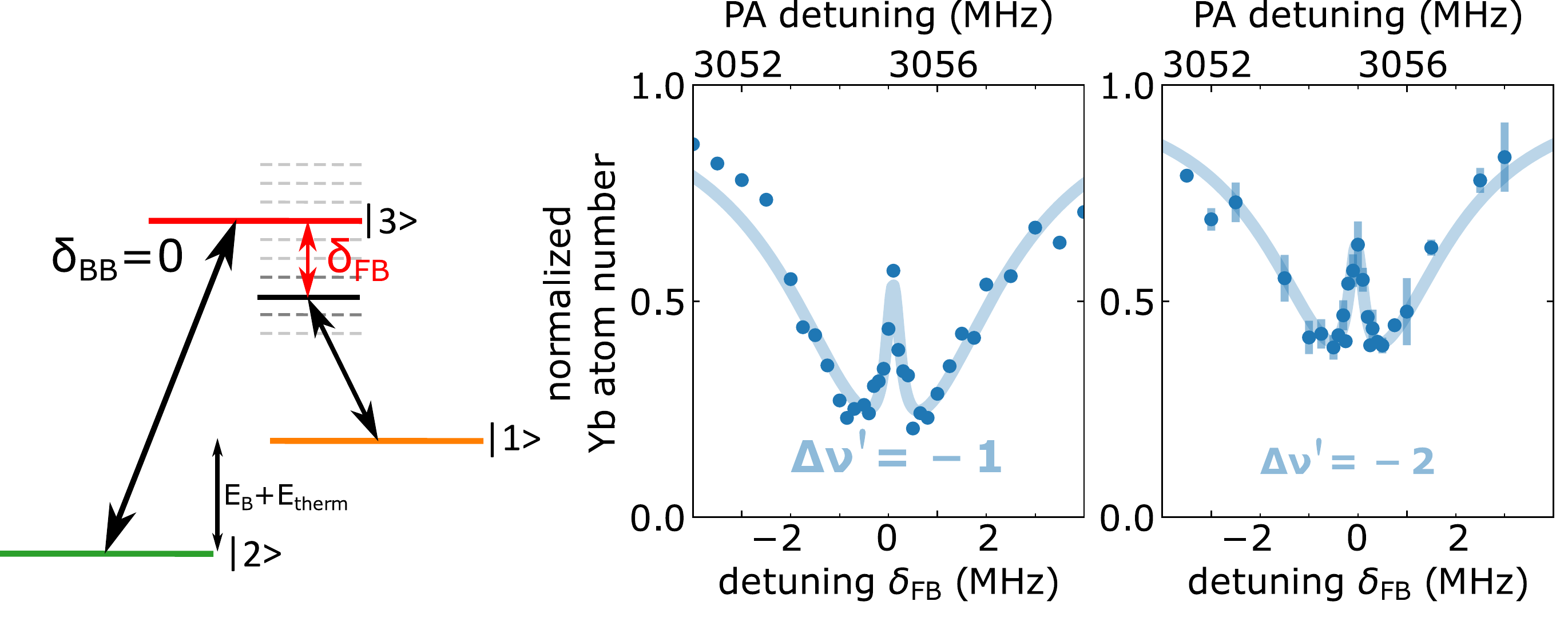}
\caption{\label{fig:2photpa} Observation of 2-photon photoassociation resonances by (a) scanning the bound bound laser with the free-bound laser locked on resonance and (b) by scanning the free-bound laser while the bound-bound laser is locked on resonance (dark-resonance configuration). The energy levels on the sketches on the left hand side are labeled with $\ket{1}$ for the scattering state, $\ket{2}$ for the bound state in the electronic ground state and $\ket{3}$ for the electronically excited state. The scattering state and the vibrational ground state are separated with the binding energy $E_{\text{B}}$ and the thermal energy $E_{\text{therm}}$.}
\end{figure}

Precise knowledge of weakly bound states in the electronic ground states will be critical for finding pathways to efficient molecule production, be it through photoassociation or magnetoassociation. The measurement of these binding energies is possible by 2-photon-photoassociation \cite{abraham95spectroscopic}, where a second laser field couples the bound state in the electronically excited state to a bound state in the electronic ground state.
Binding energies for the two least bound states in the electronic ground state of \textsuperscript{87}Rb\textsuperscript{170}Yb have previously been determined by 2-photon photoassociation near the Rb D1-line with uncertainties of around $15\,\mathrm{MHz}$ \cite{munchow2011two,Borkowski2013rbyb}. We label the vibrational states with $\Delta \nu ^\prime$ counting from the threshold with $\Delta \nu ^\prime = -1$ being the least bound state. Here we report an improvement of two orders of magnitude by performing 2-photon-photoassociation near the intercombination line.

By keeping the free-bound laser stabilized to one of the free-bound transitions\footnote{The data presented here is obtained using transition \textit{A}, but we have verified similar spectra can be obtained using transition \textit{B}.} ($\delta_\mathrm{FB} = 0$) and scanning the laser driving the bound-bound transition (bound-bound laser) and thus the 2-photon detuning $\Delta_{2\gamma}$, the photoassociation loss is suppressed when the bound-bound laser comes close to resonance, as shown in Fig.~\ref{fig:2photpa}a. The 2-photon detuning is given by $\Delta_{2\gamma} = \delta_\mathrm{FB} - \delta_\mathrm{BB}$ and its magnitude is equal to that of the scanning photon because we always keep one transition on resonance.
On the other hand, when keeping the bound-bound laser on resonance and scanning the frequency of the free-bound laser we obtain the well known dark resonance lineshape \cite{fleischhauer2005electromagnetically} shown in  Fig.~\ref{fig:2photpa}b.

In all the experiments presented here, the atoms are only weakly trapped with $\hbar\omega_\mathrm{trap} \ll k_BT$ and the scattering state is thus a continuum with thermally distributed energies rather than a well defined single state. The mean collision energy is on the order of $h \cdot 0.2 \,\mathrm{MHz}$ (ensuring that only s-wave collisions are relevant) for typical experimental conditions, . This gives rise to an asymmetrical broadening of the observed lines that is obvious in the spectra in  Fig.~\ref{fig:2photpa}a and somewhat more subtle in the dark resonances in Fig.~\ref{fig:2photpa}b. In the latter case it also prevents us from extracting the true lifetime of the dark state -- which is expected to be much longer than that corresponding to the thermal linewidth -- from the measured spectra. 
This thermal spread of energies poses a significant obstacle to the implementation of coherent techniques like STIRAP. To implement a STIRAP scheme for the production of RbYb molecules, this will have to be overcome in future experiments, for example by tight confinement of the atoms in an optical lattice. 

 All spectra are fitted using a thermal average over an expression for the loss rate obtained from solving the optical Bloch equations, which is given by
 \cite{fleischhauer2005electromagnetically}
\begin{equation}\label{eq:2photpa}
\frac{\dot N }{N}=
\frac{\Omega_\mathrm{FB}^2 \left [ 4\Gamma \Delta_{2\gamma}^2 + \Gamma_\mathrm{eff}\left ( \Omega_\mathrm{BB}^2 + \Gamma_\mathrm{eff}\Gamma\right ) \right ]}{\left |\Omega_\mathrm{BB}^2 +  \left ( \Gamma + 2i\delta_\mathrm{FB}\right ) \left (  \Gamma_\mathrm{eff} + 2i\Delta_{2\gamma}\right ) \right | ^2}
\end{equation}
with the free-bound and bound-bound Rabi frequencies $\Omega_\mathrm{FB}$ and $\Omega_\mathrm{BB}$, the decay rate of the excited state $\Gamma$ and the effective decay rate $\Gamma_\mathrm{eff}$ accounting for decoherence of the dark state.
 For the data shown, the values extracted from the fit are $\Omega_\mathrm{FB} \approx 2\,\pi\cdot 1 \,\mathrm{kHz}$ and $\Omega_\mathrm{BB} \approx 2\,\pi\cdot 1 \,\mathrm{MHz}$ for both transitions, with a free-bound laser intensity of $26\,\mathrm{W/cm^{2}}$ and bound-bound laser intensities of $5\,\mathrm{W/cm^{2}}$ for the $\Delta\nu^\prime = -1$ state and $1\,\mathrm{W/cm^{2}}$ for the $\Delta\nu^\prime = -2$ state, respectively.
 
For both observed states the Zeeman shift of the 2-photon transition was determined to be less than $0.1\,\mathrm{MHz/G}$, consistent with zero. It can thus be assumed that the Zeeman levels probed in the ground states are of the same $(F=1,~ m_F=-1)$-character as the scattering state.
The fact that no transitions with a different magnetic moment were observed indicates that the bound-bound laser does not induce a significant coupling between the Rb Zeeman or hyperfine states. 

Taking into account the thermal shifts, the measured binding energies are $E_B(\Delta\nu^\prime = -1) = h\cdot101.9(1)\,\mathrm{MHz}$ and $E_B(\Delta\nu^\prime = -2) = h\cdot1011.0(1)\,\mathrm{MHz}$. Future work will extend these measurements to more deeply bound levels, in particular those with binding energies just above the hyperfine splitting of Rb, which give rise to magnetic Feshbach resonances at moderate fields. For ${}^{87}$Rb${}^{170}$Yb, this would be the $\Delta\nu^\prime = -4$ state with the lowest resonance predicted around 1300\,G \cite{brue13prospects}.

\section{Conclusion}
 We have observed a pair of photoassociation resonances in \textsuperscript{87}Rb\textsuperscript{170}Yb and characterized them with regards to photoassociation rate, binding energy and Zeeman shift. 
To the best of our knowledge this presents the first observation of intercombination line photoassociation in RbYb. 
We have identified possible causes for the failure to observe more resonances, but like the assignment of the observed lines this question will require further spectroscopy and quantum chemistry calculations well beyond the scope of this work.

As expected, the observed PA rates are orders of magnitude smaller than those observed near the alkali D-lines. However, by preparing both species in a mixed Mott insulator the achievable Rabi frequencies may be greatly enhanced in future experiments. Starting from the motional ground state of a lattice site rather than a thermal continuum will enable much greater control over the association process and may even allow for free-bound STIRAP \cite{ciamei2017efficient}.
However, due to the unfavorable ratio of PA rate to off-resonant atomic excitation on the observed lines, it may be preferable to perform photoassociation near the D1 line of rubidium.

Nonetheless, intercombination line photoassociation is a powerful spectroscopic tool. 
We have demonstrated its use for the precise measurement of bound state energies by 2-photon photoassociation, improving the uncertainties on previously determined values by two orders of magnitude.
The precise knowledge of bound state energies in the electronic ground state will allow us to pinpoint the predicted position of magnetic Feshbach resonances, which may provide an alternative route for the initial association of weakly bound molecules. 
This spectroscopy technique will also allow for highly sensitive detection of light shifts to the bound levels. This will be invaluable for spectroscopy of more deeply bound states in the electronically excited potential, which will be required to identify a suitable intermediate level for STIRAP transfer to the absolute ground state \cite{winkler2007coherent}.

\begin{acknowledgments}

We thank Ralf Stephan for his work on the electronic and mechanical components and Matthew D. Frye for insightful discussions. T.F. acknowledges a fellowship from Prof.-W.-Behmenburg-Schenkung. The experimental apparatus was funded by DFG under grant number INST 208/614-1 FUGG.

\end{acknowledgments}

\appendix
\section{Laser stabilization}\label{sec:laserstab}
The free-bound photoassociation laser is stabilized at a tuneable offset of up to $11\,\mathrm{GHz}$ to the MOT laser using a beatnote lock. The MOT laser in turn is stabilized to a Zerodur cavity over short timescales and to the atomic transition via active stabilization of the vertical MOT position \cite{sillus2021active} over long timescales. The beatnote utilizes the fundamental output of fiber lasers at 1112\,nm before frequency doubling to 556\,nm, effectively doubling the accessible offset frequency range.

For beatnote detection we utilize a so called small formfactor pluggable (SFP) module intended for fiberoptic networking equipment operating at 1.3 or $1.5\,\mu \mathrm m$. This module provides an InGaAs detector coupled to a single mode fiber socket, a transimpedance amplifier and a limiting amplifier in a compact package and requires only minimal supporting circuitry, while providing bandwidth and signal-to-noise far exceeding that of a simple photodiode connected to a 50$\,\Upomega$ RF amplifier. 

To lock the beatnote to the desired offset frequency we use an \textit{Analog Devices ADF4159} fractional-N phase-locked loop (PLL) chip mounted on an evaluation board. This allows us to lock the beatnote to arbitrary offset frequencies up to $\sim 10\,\mathrm{GHz}$ without the need for a matching microwave reference frequency.
We note the range of offset frequencies explored is not limited by the locking setup, but rather by the time consuming nature of the experiments.

\section{Bound-bound light generation}\label{sec:boundbound}

Light used to drive the bound-bound transitions in \autoref{sec:2photpa} is derived from the same laser used for the free-bound transition. To bridge the frequency gap, two different methods are used. 

For the $\Delta \nu^\prime = -1$ transition the frequency difference is realized by frequency shifting in two acousto-optical modulators (AOM), before both beams are combined and coupled into a common single-mode fiber to ensure overlap.

The $\Delta \nu^\prime = -2$ transition on the other hand is driven by a sideband generated using a bulk electro-optical modulator (EOM) placed in the free bound laser path.
A split ring resonator \cite{Kelly87efficient} allows us to obtain a modulation index of $\beta \approx 1$, corresponding to 20\,\% of the input power in each 1st order sideband, at a moderate drive power of $1\,\mathrm{W}$. As the resonator can be constructed and tuned by simple means, this provides an attractive way for generating light at offset frequencies in the GHz range, where typical AOM setups become impractical. However, it comes at the cost of having both the carrier and further, undesired sidebands present in the output. In particular, the third harmonic of the $\Delta\nu^\prime = -2$ transition probed in Sec.~\ref{sec:2photpa} is very close to the binding energy of the intermediate states used here. This leads to one such sideband being almost resonant with the atomic transition and causing excessive trap loss even at low powers.
\typeout{}
\bibliography{RbYbPA}
\end{document}